\documentstyle[psfig]{elsart}
\begin{document}
\begin{frontmatter}
%%%%%%%%%%% PAPER %%%%%%%%%%%%%%%%%%%
\title{On the macroscopic limit of nuclear collective
motion and its relation to chaotic behavior}
\author{H.~Hofmann} 
\thanks{Supported in part by the Deutsche Forschungsgemeinschaft}
\thanks{I will be reporting mainly on topics developed together
with F.A. Ivanyuk, D. Kiderlen, A.G. Magner and S. Yamaji,
whose collaboration is gratefully acknowledged, but the whole
responsibility for this text is on my side. If not
stated otherwise, details may be found in my article in
Phys. Rep. 284 (4\&5) (1997) 137-380, for shorter reviews see
also nucl-th/9703056 and /9809017.
}
\address{Physik-Department, TU M\"unchen, D-85747 Garching\\
e-mail: hhofmann@physik.tu-muenchen.de\\
http://www.physik.tu-muenchen.de/lehrstuehle/T36/e/hofmann.html}
\thanks{Prepared for the "V. Workshop on non-equilibrium at
short time scales", \\Rostock, April 1998}

\end{frontmatter}
%%%%%%%%%%%%%%%%%%%%%%%%%%%%%%%%%%%%%%%%%%%%%%%%%%%%%%%%%%%%%%%%%%%
In this note we concentrate on slow collective motion of
isoscalar type at small but finite excitations, as given in
nuclear fission, for instance. At the end we are going to
examine how chaotic behavior might influence typical transport
properties of nuclear dynamics. The essential features may be
examined by introducing (a) collective variable(s). Typically,
the time scale of such a $Q$ is longer by about a factor of 5 to
10 than the one for the dynamics of the residual, "nucleonic"
degrees of freedom. Then it is fair to assume that the latter
are in a local equilibrium, the properties of which vary with
$Q$. A tractable but realistic description of nucleonic motion
may be based on the deformed shell model, which, in combination to the
Strutinsky procedure, is known to allow one calculating the
static energy as function of $Q$ and $T$ to satisfactory precision.

{{\bf A locally harmonic approximation:}}
For the model just explained the Hamiltonian $\hat H(\hat
x_i,\hat p_i,Q)$ will depend on $Q$ in parametric way.
To understand the dynamics in the neighborhood of some $Q=Q_0$
one may expand to second order (thus establishing a {\it locally
harmonic approximation}) 
\begin{equation}
\hat H(Q(t)) \simeq \hat H(Q_0)  + 
     (Q(t)-Q_0){\hat F} +
     {1\over 2}(Q(t)-Q_0)^2 \left<{\partial^2\hat H\over \partial Q^2}(Q_0)
    \right>^{\rm qs}_{Q_0,T_0}\, .
\end{equation}
The term of second order is treated on quasi-static
average specified by the $\hat H(Q_0)$.
The coupling between the collective and the nucleonic
degrees of freedom is given by the term in the middle. Its form allows for
an application of {\it linear response theory} to get for the
average induced force $ \delta
 \langle \hat F \rangle _t = -\int_{-\infty}^\infty
\,ds\,{\widetilde{\chi}}(s) \bigl(Q(s) -Q_0\bigr)$. 

{{\bf Local self-consistency:}}
To probe the behavior of {\it all} degrees of freedom
one may introduce a time dependent
"external field" $f_{\rm ext}(t)$ by adding  $f_{\rm ext}(t)
\hat F$ to the Hamiltonian (1).  For the {\it "collective"
response function} $ \chi_{{\rm coll}}(\omega)$ one gets
\begin{equation}
  \chi_{{\rm coll}}(\omega) = {\chi(\omega)\over 1+k\chi(\omega)}
\qquad {\rm with} \qquad
 \delta  \langle \hat F \rangle _\omega   = - \chi_{{\rm coll}}(\omega) f_{\rm 
ext}(\omega)   \end{equation}
Within an {\it "adiabatic picture"}, the coupling constant $k$ is given by
\begin{equation}
-k^{-1} = \left<{\partial^2\hat H\over \partial Q^2}(Q_0)
    \right>^{\rm qs}_{Q_0,T_0} + (\chi(0)-\chi^{\rm ad})
 = \left.{\partial^{2}E(Q,S_{0})\over \partial  Q^{2:}}\right\vert_{Q_{0}}
 +\chi (0)    \end{equation}
with $\chi^{\rm ad}$ being the adiabatic susceptibility
and $E(Q,S_{0})$ the internal energy at given entropy $S$.
Different both to the case of $T=0$ as well as to the "diabatic
picture" --- corresponding to common RPA --- the $k$ depends
sizably on $T$.

{{\bf Microscopic origin of irreversibility:}}
Already at quite small excitations a nucleus must be considered
an open system, as it may decay by emission of $\gamma$'s,
nucleons etc. Thus each level has a natural finite width.  For
any model state, this width is enlarged considerably by residual
interactions $\hat V^{(2)}_{\rm res}$  which couple the "simple"
states to "more complicated" ones. Because of the large increase
of the density of levels this mechanism gets effective already
at low temperatures. On the level of single particle motion the
effects of the $\hat V^{(2)}_{\rm res}({\hat x}_i,{\hat p}_i)$
may be parameterized through a complex self-energy
${\Sigma}(\omega\pm i \epsilon,T) = \Sigma^\prime(\omega,T) \mp
(i/ 2) \Gamma(\omega,T) $.  The intrinsic response function
$\chi(\omega)$ is calculated after replacing the single particle
strength $\varrho_k(\omega)= 2\pi\;\delta(\hbar \omega - e_k) $
by $ \varrho_k = \Gamma(\omega) /  \left((\hbar \omega -e_k
-\Sigma^\prime(\omega))^2 +   \left({\Gamma(\omega)/ 2}
\right)^2\right)$ with
\begin{equation}
\Gamma=
    {1\over \Gamma_0}\;{(\hbar \omega - \mu)^2 + \pi^2 T^2 \over 
    1 + \left[(\hbar \omega - \mu)^2 + \pi^2 T^2 \right] /c^2}
     \end{equation}
and $\mu$ being the chemical potential. The $\Gamma$ may be said
to represent "collisional damping". In numerical computations,
the following values have mostly been used for the parameters
entering here: $\Gamma_0 = 33\;{\rm MeV} $ and $ c= 20 \;{\rm
MeV}$. The width $\Gamma(\omega,T)$ takes on sizable values
already at moderate excitations $\hbar \omega - \mu$ above the
Fermi surface $\mu$. For such reasons mean field approximations
do not appear adequate at finite thermal excitations.

{{\bf Damped collective modes:}}
The microscopic damping mechanism forces collective
motion to be damped, as may be inferred from the collective
strength distribution given by $\chi_{{\rm
coll}}^{\prime\prime}(\omega)$. Take  
some individual peak of the latter located around  $\omega_1$
to represent a collective mode. This peak may be approximated by
the response function of a damped oscillator according to
$ [\chi_{{\rm coll}}(\omega)]^{-1} \simeq
[\chi_{\rm osc}(\omega)]^{-1}=  - \omega^2 M(\omega_1)-i
\omega \gamma(\omega_1) + C(\omega_1)$. defined by
the transport coefficients of average motion:
inertia $M$, friction $\gamma$ and local stiffness $C$.
The transfer of collective energy into "heat"
of the nucleonic degrees of freedom is determined by (with
$q=Q-Q_0$) 
\begin{equation}
-{d\over dt} E_{\rm coll}  \equiv
    -{d\over dt}\left({M(\omega_{1})\over 2}{\dot q}^{2}
    +{C(\omega_{1})\over 2}q^{2}\right) = \gamma(\omega_{1}){\dot q}^{2}
    \equiv  T {d\over dt}S  \end{equation}
Often the friction coefficient may well be approximated by the
following form of the so called "zero frequency limit"
\begin{equation}
\gamma\approx \gamma(0) = \int_{-\infty}^\infty \,ds\,{\widetilde{\chi}}(s)s
  = -i \left({\partial \chi(\omega) \over \partial \omega}\right)_{\omega =0}
   \end{equation}

{{\bf The macroscopic limit:}}
Many models describe the nucleus as a macroscopic system,
similar to a drop of nuclear liquid. Such a picture may be expected to
represent a nucleus realistically at larger $T$ where shell effects
become less effective --- as it is well known from studies  of the
static energy. With respect to the dependence on $T$ our theory leads
to the following behavior:  
\begin{itemize}
\item{} 
Above $T\approx 1.5 \cdots 2 \;{\rm MeV}$ the inertia $M$ drops to
values close to those of irrotational flow. This is
largely a consequence of the $T$-dependent coupling
constant (3) in the "adiabatic picture", which implies 
the strength distribution to concentrate in a strongly damped mode
at very low frequencies.
 \item{} At larger $T$ friction $\gamma$ reaches values in the
range of the "wall formula" $\gamma_{w.f.}$, provided that we use the
single particle width as given by (4).
 \item{} Neglecting in the $\Gamma$ of (4) both the $\omega$-dependence as
well as the "cut off" $c$, the $\gamma$ would drop 
like $T^{-2}$ for larger $T$, in {\it this sense} exhibiting features of
two-body viscosity as the dynamics becomes "dominated by collisions".
\end{itemize}
To some extent such features are found also in a model in which
a collision term is added to the Landau-Vlasov equation, but
where (following the "Kiev school") the latter is solved for
appropriate boundary conditions specified by the collective
mode; see a forthcoming paper with A.G. Magner.

{{\bf The interplay of one- and two-body viscosity:}} The behavior at
large $T$ should not disguise the fact that for the nucleus, as a
small and self-bound Fermi system, the mean field cannot be
discarded totally. On the other hand, and as indicated earlier,
to us it does not make much sense either, to neglect 
the coupling to more complicated states. One may ask the
question whether or not it may be possible to disentangle the
two components of friction. Inspecting the microscopic formulas
such a conjecture does not appear meaningful. The main reason is
found in the fact that the effects of the mean field, as
visualized through the one body operator $\hat F$, and of the
residual interaction are interwoven in a highly non-linear way,
as may be read off directly from the structure of the collective
response function (2). At very small excitations, on the other
hand, say at $T=0$, nuclear friction is bound to become very
small if not exactly zero. This is due to the presence of a gap
in the low energetic spectrum of nucleonic excitations. For an
inclusion of pair correlations into the present formulation see
a forthcoming paper with F.A. Ivanyuk.

{{\bf The wall formula:}} 
It has been demonstrated in various ways that and how the $\gamma(0)$
of (6) turns into the $\gamma_{w.f.}$ of the "wall formula". Generally
speaking, the latter is reached as a macroscopic limit of the  $\gamma(0)$
if evaluated within the picture of {\it pure independent particle
motion}, i.e. for $\hat V^{(2)}_{\rm res} =0$ or $\Gamma=0$.
This limit may be defined literally by letting the
size of the system become infinitely large or by
applying Strutinsky smoothing procedures. In the former case
collective dissipation may indeed arise from collisions of the particles with
the moving "wall", with irreversibility showing up in a kind of
"thermodynamic limit". Recently, a Strutinsky smoothing has been
applied to the model where a finite system of nuclear dimensions, but
consisting of independent particles, is forced to undergo shape 
vibrations, following some oscillating external source. Treating the
latter within time dependent perturbation theory, the excitation left
in the system after one cycle of the external field shows the typical
strength function behavior, as function of the vibrational frequency.
Applying a Strutinsky smoothing to these strength functions it is seen
that the averaged energy behaves like expected from wall friction; for
details see nucl-th/9709043. It is important to note that
typically such smoothing procedures involve averaging over
intervals in energy of the order of $10\,-\,20 \;{\rm MeV}$.
This fact allows one to understand physically why the wall
formula does not reflect any shell effects, and, hence, that it
is insensitive to changes in $T$.

{{\bf Heat pole:}} The (dissipative part of the) response function is
associated to the correlation function by the fluctuation dissipation
theorem (FDT) $\psi^{\prime \prime}(\omega) = \hbar \coth({\hbar
\omega \over 2T})\, \chi^{\prime \prime }(\omega) $.  
In case that the intrinsic states have zero width the
$\psi^{\prime \prime}(\omega)$ has the following structure 
$\psi^{\prime \prime}(\omega)=\psi^{0} 2\pi \delta(\omega)
 \; + \;_R\psi^{\prime \prime}(\omega)$ 
with $\psi^{0} = T\Bigl( \chi^{\rm T} - \chi(0) \Bigr)$
and $\chi^{\rm T}$ being the isothermal susceptibility.
In analogy to transport in infinite systems the contribution
at $\omega=0$ may be called the "heat pole". Due to the damping
mechanism mentioned above a smooth peak develops 
\begin{equation}
 {_0\psi^{\prime \prime}}(\omega)= \psi^0 2\pi \delta(\omega) \qquad
    \Longrightarrow \qquad {_0\psi^{\prime \prime}}(\omega) = \psi^0 {\hbar 
\Gamma_T
    \over (\hbar \omega)^2 +
   \Gamma_T^2 /4}    
\end{equation}
Its width may estimated to be about twice the
single particle width at the Fermi surface $\Gamma_T \approx 2
\Gamma(\mu,T)$ (which in turn is close to $2T$).
According to (6) the heat pole contributes to
friction by the amount
${_0\gamma(0)} \equiv 2\hbar \psi^{0}/ (T \Gamma_T) $.

{{\bf Nuclear ergodicity:}} The residue of the heat pole reflects
ergodic properties, which may become apparent after introducing
the adiabatic susceptibility:
$\psi^{0}/T = \chi^{\rm T} - \chi(0)  =
    \Bigl( \chi^{\rm T} - \chi^{\rm ad} \Bigr)-
    \Bigl( \chi^{\rm ad} - \chi(0) \Bigr) $.
For the nuclear case the difference $\chi^{\rm T} -\chi^{\rm
ad}$ can be shown to be small. The difference $\chi^{\rm ad}
-\chi(0)$, on the other hand, is known to vanish if two
conditions are met: The system should have no degeneracies and
the spread in the energy of the states contributing should be
sufficiently narrow. The latter property may cause problems for
the nuclear case if the canonical distribution must be applied.
The first property definitely will do so for the model of
independent particles, but {\it adequate inclusion of $\hat
V^{(2)}_{\rm res}$ can be expected to supply sufficient level
repulsion}. Indeed, an application of the {\it Random Matrix
Model to the linear response approach} is seen to simulate
ergodic behavior. However, it must be said that our present way
of handling the $\hat V^{(2)}_{\rm res}$ through self-energies
is not good enough as by way of the form (4) degeneracies are
not lifted. This is unfortunate, as otherwise our formulation
allows one to include the many important features of collective
motion mentioned above, which as yet, and to the best of my
knowledge, are not amenable by applications of the RMM, in
particular not at smaller excitations.

{{\bf Traces of chaotic motion:}} Nuclear dynamics is known to
exhibit chaotic features. The most prominent one is that given
by {\it Wigner's law for the distribution of levels of the
compound nucleus}, as seen in neutron resonances, for instance,
thus being associated to an excitation of the order of the
neutron binding energy. This property is {\it intimately
connected to the existence of the residual interactions $\hat
V^{(2)}_{\rm res}$} mentioned before, which {\it couple simple
configurations to "more complicated ones"}. The simple ones may
be understood of being those of the {\it bare shell model}, viz
those of the mean field approximation, which would lead to the
{\it Poisson distribution not seen in experiment} (in the range
of excitations mentioned before). In addition, for a typical
mean field the {\it motion of a nucleon itself may already show
some chaotic behavior}. This can be expected to be the more so
the more this field is deformed; but notice that this fact does
not rule out the appearance of shell effects (and thus of
regular behavior) at finite deformations. Certainly, the study
of such phenomena in their own are of considerable interest. In
{\it connection to transport theory} however, it may be vital to
{\it understand in which way chaotic dynamics is related to
dissipative behavior}. Without any doubt, such a relation
exists, indeed, in as much as $\hat V^{(2)}_{\rm res}$ is
involved; please recall the remarks made in the previous
section.  However, there are claims that the {\it "wall formula"
can be justified on the basis of nucleonic motion within a
sufficiently complex potential}, discarding any $\hat
V^{(2)}_{\rm res}$. In our understanding this statement is {\it
not justified}, for two reasons: (i) At finite $T$, the $\hat
V^{(2)}_{\rm res}$ {\it must not} be neglected, according to (4)
its influence even increases with $T$. (ii) The very form of the
wall formula --- if applied to independent particle motion ---
does involve averaging over large intervals which in themselves
wipe out any other microscopic details, see again
nucl-th/9709043. In addition it may be stated that in our
microscopic computations hardly any traces have been seen that
friction would become bigger for more complex shapes.

\end{document}